\documentclass{article}
\pdfoutput=1

\usepackage{arxiv}

\usepackage{amssymb}
\usepackage{amsthm}
\usepackage{amsmath}
\usepackage{xcolor}
\usepackage{soul}
\usepackage{multirow}
\usepackage{subcaption} 

\usepackage[utf8]{inputenc} 
\usepackage[T1]{fontenc}    
\usepackage{hyperref}       
\usepackage{url}            
\usepackage{booktabs}       
\usepackage{amsfonts}       
\usepackage{nicefrac}       
\usepackage{microtype}      
\usepackage{graphicx}
\usepackage[numbers]{natbib}
\usepackage{doi} 
\graphicspath{ {./figures/} }

\title{Rapid Production of Accurate Embedded-Atom Method Potentials for Metal Alloys}

\author{
Elan J. Weiss \\
Department of Materials Science and Engineering\\
The Ohio State University \\
Columbus, OH 43201
\texttt{weiss.443@osu.edu} \\
\And
Logan Ward \\
Department of Materials Science and Engineering\\
The Ohio State University \\
Columbus, OH 43201
\texttt{weiss.443@osu.edu} \\
\And
Christian Oberdorfer \\
Department of Materials Science and Engineering\\
The Ohio State University \\
Columbus, OH 43201
\texttt{weiss.443@osu.edu} \\
\And
Travis Withrow \\
Department of Materials Science and Engineering\\
The Ohio State University \\
Columbus, OH 43201
\texttt{weiss.443@osu.edu} \\
\And
David C. Riegner \\
Department of Materials Science and Engineering\\
The Ohio State University \\
Columbus, OH 43201
\texttt{weiss.443@osu.edu} \\
\And
Anupriya Agrawal \\
Department of Materials Science and Engineering\\
The Ohio State University \\
Columbus, OH 43201
\texttt{weiss.443@osu.edu} \\
\And
Wolfgang Windl \\
Department of Materials Science and Engineering\\
The Ohio State University \\
Columbus, OH 43201
\texttt{windl.1@osu.edu} \\
}

\begin{document}
\maketitle
\begin{abstract}
A critical limitation to the wide-scale use of classical molecular dynamics for alloy design is the limited availability of suitable interatomic potentials. Here, we introduce the ``Rapid Alloy Method for Producing Accurate General Empirical Potentials” or RAMPAGE, a computationally economical procedure to generate binary embedded-atom model potentials from already-existing single-element potentials that can be further combined into multi-component alloy potentials. We present the quality of RAMPAGE calibrated Finnis-Sinclair type EAM potentials using binary Ag-Al and ternary Ag-Au-Cu as case studies. We demonstrate that RAMPAGE potentials can reproduce bulk properties and forces with greater accuracy than that of other alloy potentials. In some simulations, it is observed the quality of the optimized cross interactions can exceed that of the original `off-the-shelf' elemental potential inputs.
\end{abstract}



\section{Introduction}
\label{sec:intro}

Classical molecular dynamics (MD) simulations have become ubiquitous tools for simulating complex behavior in materials systems on the atomic level for increasingly larger systems. Classical MD simulations are based on  interatomic potentials, which approximate complex interatomic interactions using computationally efficient functions in analytical or tabulated form. These functions range from simple pair interactions to more complex formulations involving a gamut of other terms, e.g.\  bond-order or embedding terms that correct the energy of an atom based on its coordination, three-body terms that stabilize bond angles, or four-atom terms that control torsion. More recent work has proposed to substitute these physically motivated functions by data-driven machine learning formulations. MD simulations can achieve simulation sizes and times orders of magnitude larger than more rigorous \textit{ab initio} methods such as density functional theory (DFT) due to the fast energy and force evaluation paired with linear $O(N)$ scaling of computational cost with respect to number of particles $N$, reaching on modern GPU implementations a performance of 100 million atom timesteps per node and second~\citep{sikandar_mashayak_lammps_nodate}. 

Despite the computational advantage of using classical MD, the limited availability of suitable interatomic potentials (IAPs) remains a substantial barrier to its widespread deployment. The NIST Interatomic Potential Repository~\citep{becker_considerations_2013,hale_evaluating_2018}, now integrated with OpenKIM~\citep{tadmor2011potential}, represents one of the most reliable databases of interatomic potentials for use with MD. Of the 28 transition metal elements (excluding Hg and all heavier elements), there are 24 available elemental potentials more advanced than a Lennard-Jones model. When considering multi-component potentials, there are 157 available binaries out of a possible 378, and 85 ternaries out of a possible 3,376. This trend of limited multi-component potentials continues into 4 and 5 component potentials with 47 out of 20,475 and 15 out of 98,280 available potentials respectively~\citep{becker_considerations_2013,hale_evaluating_2018,tadmor2011potential}. As engineered alloys may contain over ten elements, the limited availability of multi-component interatomic potentials continues to pose a substantial bottleneck~\citep{gardner_use_2005,ryan_why_2002}. 

This is further complicated by the necessary compromises between computational efficiency and modeling accuracy, as well as conscious choices by a potential developer to optimize an interatomic potential to a given system or property. 
Not only do different potential forms excel at modelling different materials systems, but computational costs may limit the use of some types of potentials. For example, potentials with highly complex functional forms or machine learning formulations such as, SNAP~\citep{thompson2015spectral}, ReaxFF~\citep{chenoweth_reaxff_2008}, GAP~\citep{bartok2010gaussian}, and other ML potentials may have a wide degree of accuracy and applicability to many different materials classes, but their computational expense, which can be easily two to three orders of magnitude higher than traditional potentials, limits the time and length scales that can be probed~\citep{plimpton_computational_2012}. For metals, embedded atom method (EAM) potentials have been used widely because of their low computational cost and ability to accurately model bulk properties and defects for many elemental metals and alloy systems~\citep{daw_embedded-atom_1984}. Despite the simplicity of the model and more recent advancements in MD potentials, EAM potentials remain widely used by the community~\citep{plimpton_computational_2012,zhou_misfit-energy-increasing_2004,sheng_highly_2011}.

Here we introduce the Rapid Alloy Method for Producing Accurate General Empirical (RAMPAGE) potentials, with the goal of bridging the gap between available multi-component IAPs and real engineering alloys, specifically, by using published elemental potentials as model inputs and fitting only the needed cross interaction terms. While an even simpler way has been previously proposed to produce alloy potentials, the Johnson Alloy Model (JAM) \citep{zhou_misfit-energy-increasing_2004,sheng_highly_2011,johnson_alloy_1989},  where the cross terms are averaged from the elemental potentials, the lack of control over the enthalpy of mixing through a separate term is often too restrictive to reproduce observed alloying and compound formation behavior as we will discuss below.  

In the next section, we introduce the RAMPAGE framework where a separate cross term is introduced to overcome the JAM shortcomings. We then discuss two case studies of fitted Finnis-Sinclair RAMPAGE potentials for systems with starkly contrasting alloying behavior, a binary Ag-Al potential and a ternary Ag-Au-Cu potential and benchmark them against the corresponding JAM potentials. The fitted RAMPAGE potentials are then benchmarked against DFT data for a problem outside of the fitting input set (which are all based on bulk systems) to gauge their transferability, for which we chose desorption from a (100) surface. This is an important problem requiring fast potentials, since recent advances in combining a finite-element Poisson solver with an MD code have shown that field-evaporation from a realistic-size atom-probe tomography (APT) sample allows to understand, analyze and uncertainty-quantify APT results to an unprecedented degree, taking 3D atomic-scale characterization to a completely new level~\citep{parviainen_atomistic_2015, yao_effects_2015, oberdorfer_influence_2018, qi_ab-initio_2022}. 

\section{Embedded Atom Method}
\label{sec:EAM}

The embedded atom method was first introduced by Daw and Baskes in 1984~\citep{daw_embedded-atom_1984, daw_semiempirical_1983}, and shortly after that Finnis and Sinclair~\citep{finnis_simple_1984} proposed a similar model composed of the same constituent components. These models expanded on the earlier pair potential (PP) approach,
\begin{equation}
    \label{eq:PP}
    E_{\rm PP} =     \frac{1}{2}\sum_{i \neq j} \phi_{\alpha \beta}\left( r_{ij} \right),
\end{equation}
where $\phi_{\alpha \beta}\left( r_{ij} \right)$ is the interaction or bond energy as a function of distance $r$ for all atom pairs $ij$ in the system and $\alpha$ ($\beta$) denotes the element type of atom $i$ ($j$), 
by correcting it with the help of a many-body correction term $F$, the embedding function, which depends on pairwise density functions $\rho$, 
\begin{equation}
    \label{eq:EAM}
    E_{\rm FS,EAM} =     \frac{1}{2}\sum_{i \neq j} \phi_{\alpha\beta} \left( r_{ij} \right) +\sum_i F_{\alpha}\left(\sum_{j \neq i}\rho_{ \left[\alpha \right] \beta}\left(r_{ij}\right)\right),
\end{equation}
where adding the bracketed index $\alpha$ to the electron density switches the type from EAM to Finnis  Sinclair (FS)~\citep{finnis_simple_1984}. 
In the original interpretation by Daw and Baskes, $\rho$ was thought of as an electron density function, which gives the elemental or pair-wise contribution of an atom to the surrounding electron cloud, and the embedding function $F$ described the potential energy that resulted from embedding an atom into that electron cloud~\citep{wadley_mechanisms_2001}. Later interpretations note that $\rho$ can also be thought of as an effective bond-counting function and $F$ as a bond-energy correction term, since more bonds formed by the same number of electrons would result in weaker strength per bond \citep{voter1994embedded}. This interpretation is also the basis for the functionally identical glue model~\citep{ercolessi1986glue}, where less electron ``glue'' per bond leads to weaker bonding. 

Originally, in  EAM the electron density terms are solely elemental, $\rho_\beta(r)$, assuming an atom of type $\beta$ would always dump the same amount of electron density on a neighbor at a given distance $r$ irrespective of that neighbor's type and size, thus the only cross-term is the pair potential $\phi_{AB}$. This assumption is removed within the Finnis-Sinclair (FS) formulation allowing for different functions for different types $\alpha$ and $\beta$, which introduces another cross element function, the electron density $\rho_{\alpha\neq\beta}$, besides the pair potential.

The EAM formulation relies solely on interactions between two atoms. Therefore arbitrary multi-component alloy potentials can be generated by concatenating compatible binary EAM potentials. Likewise, binary EAM potentials can be built from pre-existing elemental potentials by reusing the element specific terms and fitting only cross interaction term(s). However, as the EAM framework makes no constraints on the form of the embedding function or electron densities, and the electron density term is on an arbitrary unit axis, re-using any pre-existing potentials requires these terms be mutually compatible. This is often a natural outcome when the elemental potentials are fitted by the same author~\citep{pun_development_2009}. Otherwise, when developing binary potentials, compatibility can only be accomplished by starting from a single pre-existing elemental potential and fitting second-element and  cross-interaction term(s) based on that~\citep{williams_embedded-atom_2006}.

One of the most common methods for rapidly developing new multi-component alloy potentials is the Johnson Alloy Model  (JAM)~\citep{zhou_misfit-energy-increasing_2004,sheng_highly_2011,johnson_alloy_1989}. JAM  uses the EAM alloy formulation with $\phi_{AB}=\phi_{\alpha\neq\beta}$ as only cross term and determines it from a weighted average of the elemental pair functions,
\begin{equation}
    \label{eq:JAM}
    \phi_{AB}(r) =     \frac{1}{2} \left [ \frac{\rho_B(r)}{\rho_A(r)} \phi_A(r) + \frac{\rho_A(r)}{\rho_B(r)} \phi_B(r) \right ].
\end{equation}
The weighting factors ${\rho_B(r)}/{\rho_A(r)}$ and ${\rho_A(r)}/{\rho_B(r)}$ ensure that the resulting pair potential is well balanced; however, if one electron density is much larger than the other, one weight factor will approach zero while the other becomes very large, leading to an unbalanced pair potential. In addition, this method is still subject to the constrains of mutual compatibility, and the density function $\rho_A$ and $\rho_B$ need to be defined on ranges compatible with both embedding functions $F_A$ and $F_B$.

\begin{figure}
    \centering
         \centering
         \includegraphics[width=0.4\textwidth]{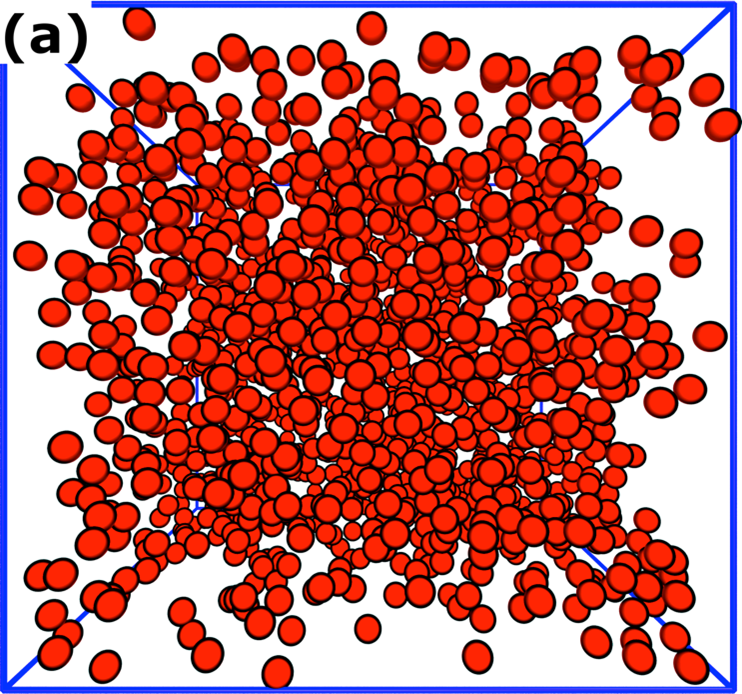}
         \label{fig:RAM_Glass}
         \centering
         \includegraphics[width=0.4\textwidth]{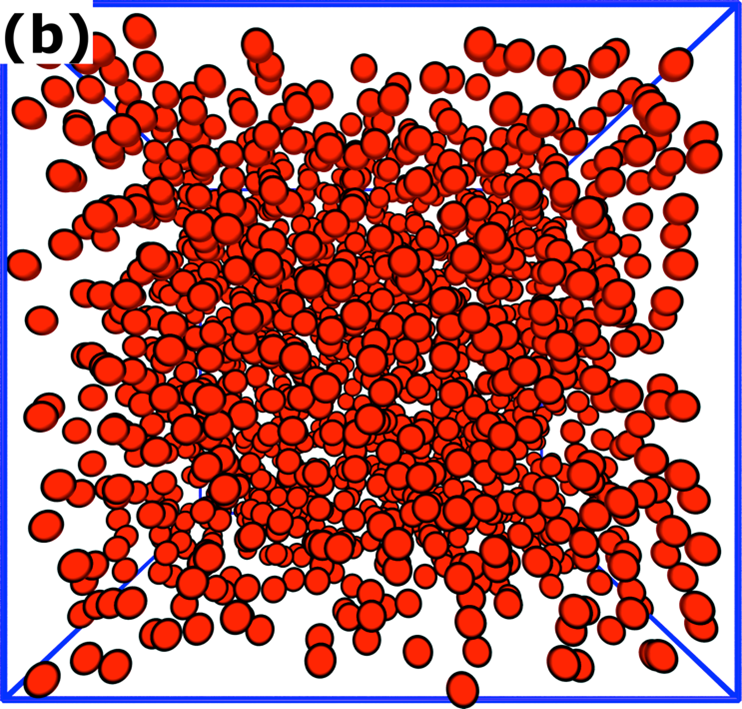}
         \label{fig:Sheng_Glass}
         \centering
         \includegraphics[width=0.4\textwidth]{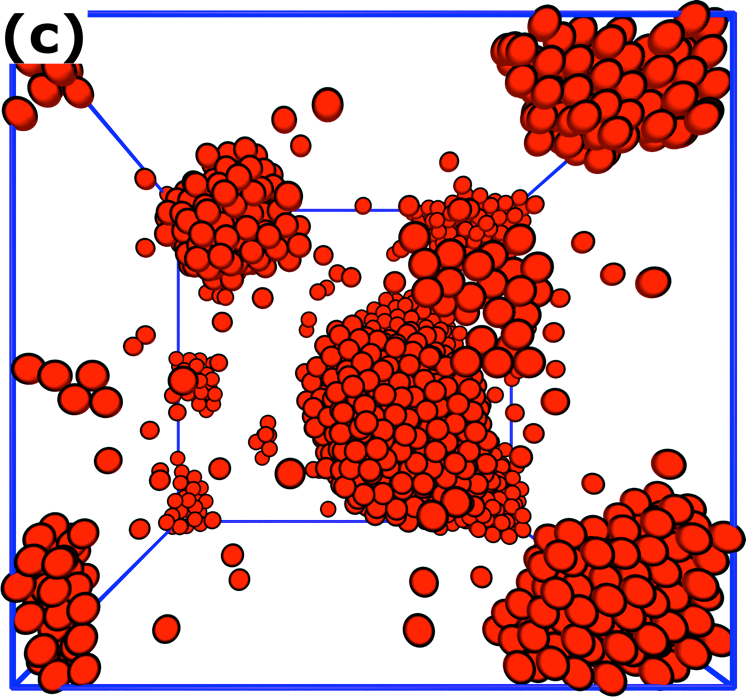}
         \label{fig:JAM_Glass}
    \caption{Aluminum distribution in Cu$_{45}$Zr$_{45}$Al$_{10}$ metallic glass at 0~K in 16000-atom models generated by simulating a rapid quench from the liquid state using potentials made by (a) RAMPAGE, (b) Sheng~\citep{cheng_atomic_2009}, and (c) the Johnson alloy model~\citep{zhou_misfit-energy-increasing_2004}.}
    \label{fig:EAMglass}
\end{figure}

This makes JAM most practical when a set of compatible elemental and binary functions is developed by the same research teams, resulting in publications of comprehensive alloys libraries. In 2001, Zhou \textit{et al.} published a database of EAM potentials for 16 metals \citep{zhou_misfit-energy-increasing_2004}, and in 2011, Sheng \textit{et al.} published a collection of 14 metal EAM potentials~\citep{sheng_highly_2011}. While JAM potentials are easily constructed, the simplicity of the model with lacking control of the mixing enthalpy can lead to predictions of inaccurate alloying behavior. As an example, Figure~\ref{fig:EAMglass} shows results for the simulated structure of a Cu$_{45}$Zr$_{45}$Al$_{10}$ metallic glass, where the vitrification has been simulated with both JAM and RAMAPGE potentials based on Sheng's elemental potentials from \citep{sheng_highly_2011}. Whereas the  RAMPAGE potential shows a uniform distribution of aluminum in agreement with experimental expectation~\citep{mendelev_development_2009}, the JAM potential predicts a distinct demixing of the aluminum.

In order to have good control over the mixing enthalpy, the potentials discussed in this work use the FS formulation. With the extra degrees of freedom provided by the cross-element densities, FS potentials have been shown to produce accurate fits in a variety of systems~\citep{mendelev_development_2009, landa_development_1998, koleske_molecular_1993, hyodo_empirical_2020}. This added accuracy and flexibility comes at no significant additional computational cost at simulation run-time.


An EAM-FS model of a binary system requires nine functions $\phi_{AA}$, $\phi_{BB}$, $\phi_{AB}$, $F_A$, $F_B$, $\rho_{AA}$, $\rho_{BB}$, $\rho_{AB}$, and $\rho_{BA}$. Of the nine functions in a binary EAM-FS potential, six describe bonding between atoms of the same element. As online interatomic potential repositories~\citep{tadmor2011potential,becker_considerations_2013,sheng_highly_2011} provide large libraries of published elemental potentials, the task of creating a binary EAM-FS potential can be simplified by using those fitted functions. This reduces the scope of the fitting problem to three remaining cross interaction terms. With these, since Eq.~\ref{eq:EAM} only contains pair interactions, multi-component FS potentials can be built by combining the constituent elemental and binary cross interaction terms. 

\section{Method}

RAMPAGE operates by creating binary EAM Finnis-Sinclair (EAM-FS) potentials from 
elemental EAM potentials from the literature, and fitting cross interaction functions to a DFT computed training set consisting of a strategically-selected set of special quasi-random (SQS) and intermetallic structures. Within RAMPAGE several analytical functions for pair potential and electron density are implemented and are selected by the user during binary fitting. The free parameters of the chosen models are then fitted by using an elitist genetic algorithm. Regardless of the chosen models, RAMPAGE binary potentials constructed using the same elemental potentials can then be assembled into multi-component interatomic potentials with no additional fitting required. 

\subsection{Standardizing Elemental Potentials}

As a consequence of using published elemental potentials, the complexity of fitting a binary EAM-FS potential is limited to three cross interaction functions; however, fitting can only be performed if the elemental potentials are compatible. There are no requirements within the EAM-FS formulation that enforce consistency across potentials. Individual authors make choices on cutoff radius, grid spacing and size. In addition, the ranges of electron density and embedding functions, which lie on arbitrary unit axes, and the form of the pair potential are all chosen based on individual reasoning. 

To ensure compatibility, the single-element potentials are initially scaled using Voter's invariant transformations
\begin{equation}
\label{eq:rhotransform}
\begin{aligned}
     & F_i \left( \rho \right) \xrightarrow{} F_i \left( \frac{\rho_i}{s} \right), \\
     & \rho_i \left( r \right) \xrightarrow{} s \cdot \rho_i \left(r \right).
\end{aligned}
\end{equation}
The embedding functions, which take the sum of local electron densities as input, are transformed to a consistent maximum $\rho$ value. In this work, a value of 400 was chosen. The electron densities are likewise scaled in order to preserve the original energetics and forces of the elemental systems using the appropriate value for $s$. 

The pair potentials are likewise scaled to have their minimum at 0.04\% of the cohesive energy using the following transformations:

\begin{equation}
\label{eq:phitransform}
\begin{aligned}
     & F_i \left( \rho \right) \xrightarrow{} F_i \left( \rho_i \right) + g \cdot \rho_i \left( r \right), \\
     & \phi_i \left( r \right) \xrightarrow{} \phi_i(r) - 2g \cdot \rho_i \left( r \right).
\end{aligned}
\end{equation}


Finally, for each case study, the largest cutoff radius among the constituent elemental potentials was applied to all functions. As EAM potentials are distributed in tabulated format with each function defined on discrete intervals, a cubic spline interpolation evaluates each component function for values not specified in the original elemental tables. 

\subsection{EAM Cross Interaction Models}

Within RAMPAGE there are several functions for pair potential and electron density implemented. A full list of the functions within RAMPAGE is detailed in the supplemental information. Both RAMPAGE potential examples discussed in this paper use the same pair potential model. The pair interaction function is fitted to a standard Morse potential \citep{voter_accurate_1986},
\begin{equation}
\label{eq:morse}
V_{AB}(r) = D \left( \exp \left( -2 \alpha \left( r - r_{\text{eq}} \right) \right) - 2 \exp \left( - \alpha \left( r - r_{\text{eq}} \right) \right) \right).
\end{equation}
The Morse function was chosen in lieu of other functions because of its simplicity. With only three fitting parameters, $D$, $\alpha$, and $r_{eq}$, Morse potentials have been shown to reliably model a number of metallic systems~\citep{voter_accurate_1986,zhou_misfit-energy-increasing_2004,sheng_highly_2011}. To ensure that the function and its first derivative approach zero at the cutoff distance $r_{cut}$, the cutoff function 
\begin{equation}
\label{eq:cutoff}
\phi_{AB}(r)=V_{AB}(r)-V_{AB}(r_{cut}) + \left(\frac{r_{cut}}{m}\right) \left[ 1 - \left( \frac{r}{r_{cut}} \right)^m \right] \left( \frac{dV_{AB}}{dr} \right)_{r=r_{cut}}
\end{equation}
was applied with $m=20$ as suggested by Voter \citep{voter1993embedded}.

Each case study was fitted using a different electron density model. For the Ag-Au-Cu potential, the electron density functions were fitted as a linear combination of the original elemental functions, where
\begin{equation}
\label{eq:interpolated}
\begin{aligned}
& \rho_{BA}(r) = \rho_{BB}(r) + S_A\left[\rho_{AA}(r)-\rho_{BB}(r)\right],\\
& \rho_{AB}(r) = \rho_{AA}(r) + S_B\left[\rho_{BB}(r)-\rho_{AA}(r)\right],
\end{aligned}
\end{equation}
with $0\le S_{A,B}\le 1$. In this form, $S_A$ and $S_B$ are fitting parameters, and the effective shape of the density clouds is assumed to be related to the elemental functions. The key advantage of this assumption is that describing the electron density now requires just two fitting parameters, and no knowledge of the original functional forms of either the electron density or embedding function to which $\rho$ acts as an input. When $0 \le S_A,S_B \le1$, equation~\ref{eq:interpolated} acts as a linear interpolation. For the potentials shown in this investigation, these values were allowed to exceed the elemental limits; however, the fitted results were within these limits for all potentials except Ag-Cu.

The interpolated electron density model produced low quality Ag-Al potentials, thus the case study presented here adopted an electron density model of the form
\begin{equation}
\label{eq:csw}
\rho(r) = \frac{1+a_1 \cos (a_0 r) + a_{2} \sin (a_0 r)}{r^b},
\end{equation}
which was proposed by Chantasiriwan and Milstein~\citep{chantasiriwan_higher-order_1996}. This model requires 4 fitting parameters per element: $a_0$, $a_1$, $a_2$, and $b$. The additional degrees of freedom associated with the larger number of parameters created fits that balanced the three properties in the training set to a far greater extent than the interpolated electron density model. 

\subsection{Density Functional Theory Training Set}
\label{sec:dftset}

For the test cases investigated here, the training set for a binary system is generated from DFT calculations of 32-atom SQS cells consisting of a 2x2x2 FCC supercell, along with B2 and L1$_2$ intermetallics. 
These calculations provide the equilibrium lattice parameters, mixing enthalpies, and bulk moduli; the three quantities of interest (QOIs) that the resulting EAM potential aims to reproduce. Here, these calculations are conducted using VASP~\citep{kresse_ab_1993,kresse_ab_1994} with projector augmented wave (PAW) pseudopotentials~\citep{blochl_projector_1994,kresse_ab_1993,kresse_ab_1994}, Perdew-Burke-Ernzerhoff (PBE) exchange-correlation functional~\citep{perdew_generalized_1996,perdew_generalized_1997}, and k-point grid spacing of 0.1~\AA$^{-1}$.

To compensate for the difference between standardized elemental potential and DFT-predicted lattice parameters and bulk moduli, a “rule-of-mixtures” approach is used. For each element, correction factors for the lattice parameters and bulk modulus are determined using
\begin{equation}
\label{eq:correctionFactors}
\begin{aligned}
c_{\rm lattice} & = \frac{1}{3} \left ( \frac{a_{\rm MD}}{a_{\rm DFT}} + \frac{b_{\rm MD}}{b_{\rm DFT}} + \frac{c_{\rm MD}}{c_{\rm DFT}} \right ), \\
c_{\rm bulk} & = \frac{K_{\rm MD}}{K_{\rm DFT}},
\end{aligned}
\end{equation}
where $a, b,$ and $c$ are the three lattice parameters and $K$ is the bulk modulus. The moduli and lattice parameters calculated for intermetallics are multiplied with an effective correction factor, which is an average of the elemental correction factors weighted by composition. For a binary system, this would correspond to

\begin{equation}
\label{eq:ruleofmixtures}
    c_{A_{1-x} B_{x}} = (1-x)c_A + x c_B,
\end{equation}
where $c_{A_{1-x} B_{x}}$ is the correction factor for a given property with elemental correction factors $c_A$ and $c_B$ and alloy stoichiometry given by $x$. This normalization of the training set data is required as elemental potentials may be fitted to experimental properties at a given temperature, \textit{ab initio} calculations at 0~K, or may not include any properties and be fitted exclusively with force matching. Thus, without normalization, the training set data may exceed the bounds imposed by a given pair of elemental potentials. This creates an intractable fitting problem that can be rectified with the normalization procedure described above. 

\subsection{Fitting Procedure}
\label{ssec:FitPrtoc}

The optimum values of all free parameters (five per Ag-Au-Cu binary, and eleven for the Ag-Al binary) are determined by minimizing the difference between the mixing enthalpy, bulk modulus, and lattice parameters for the training set structures as calculated within RAMPAGE by using the LAMMPS library interface versus corrected DFT~\citep{plimpton_fast_1995}. A genetic algorithm optimizes an objective function $\mathfrak{F}$, which we define as the weighted sum over all structures in the training set of total error in mixing enthalpy, and fractional errors in bulk modulus and lattice parameter between EAM and the corrected DFT values calculated using the relationship
\begin{equation}
\label{eq:fitnessFunction}
\frac{1}{\mathfrak{F}} = \sum_i \left[ w_H\Delta\left(\Delta H_{m,i}\right) + w_K\frac{\Delta K_i}{K_{i,DFT}} + w_a\frac{\Delta a_i}{a_{i,DFT}} \right] w_i,
\end{equation}	 	
where $\mathfrak{F}$ represents the fitness of the potential with respect to matching the corrected \textit{ab initio} values. The weight factor $w_i$ is associated with a given structure and is used in Section\ref{sec:CaseStudyAgAl} to bias a fits towards specific compositions. The weight factors $w_H$, $w_K$, and $w_a$ can in principle be freely chosen and can be used to preferentially skew behavior towards a given property in the final fit. 

Since their values however can drastically alter the behavior of the resulting potential, we have  chosen a test set-based validation method which makes the weights part of the optimization problem. Equally, for some alloys, inclusion of the intermetallic phases into the training set can have significant effect on the optimization outcome. In order to have a data-driven selection of both weights and training structures, a matrix of weights and training data sets was developed and separate fits performed for each matrix point. The performance of these potentials was then assessed against a validation set of 17 SQSs.
More details of the procedure are described in the supplemental information. In an extension of this approach, a Bayesian calibration framework was also developed to optimize and understand the limits of the parameter space, their correlations with each other, and the general performance of RAMPAGE potentials better, which can be found in Ref.~\citep{hedge_bayesian_2022}. 

\section{Case Studies}
\label{sec:CaseStudies}

The process discussed above was used within the RAMPAGE code package to fit potentials for the ternary Ag-Au-Cu and binary Ag-Al alloy systems. The Ag-Au-Cu system was selected as it is one of the few test systems that can closely approximate an ideal solid solution. This should provide the best case scenario for JAM potentials, which rely on a model that is predicated on the assumption of ideal mixing. The second case study applies RAMPAGE to fit Ag-Al binary potentials. Unlike, the first case study, Ag-Al is an immiscible system with extended low temperature solubility of aluminum in silver up to 20 atomic percent. Silver exhibits a maximum solubility of 23.5 atomic percent aluminum at 840~K, but the solubility limit rapidly becomes negligible as temperature decreases~\citep{asm_phase_2016}. This system provides a challenging test case for interatomic potential fitting in a near 'real-world' context. We benchmark the resulting RAMPAGE potentials against desorption simulations not included in the training or validation sets to show transferability of the resulting potentials. Both cases use elemental potentials fitted by Zhou et al.~\citep{zhou_misfit-energy-increasing_2004}, which allows a consistent comparison to cross interactions modeled by the JAM, for which they were originally developed. 

\subsection{Ag-Au-Cu}
\label{sec:CaseStudyAgAuCu}

The silver, gold, copper alloy system provides a close approximation of ideal behavior for benchmarking RAMPAGE potentials. This system can be well described by simple models such as JAM and even within RAMPAGE, a simple interpolated electron density model was chosen due to its ease of optimization and favorable performance. 

The procedure for creating a ternary RAMPAGE potential requires one additional step after the three constituent binaries: Ag-Au, Ag-Cu, and Au-Cu are fit. Those three binaries are concatenated into a ternary potential. 
The only modifications that need to be made to the constituent binary potentials are insuring compatible cutoff radii and grid spacing for the numeric tables. However, before the fitting of the cross terms can begin, it is important to benchmark the accuracy of the elemental potentials, as elemental interactions remain unchanged through the fitting procedure. Therefore the quality of any RAMPAGE multi-component potential will be limited by the quality of the elemental potentials selected. 


Equilibrium properties of elemental copper, gold, silver, and aluminum from both DFT and MD are shown in Table \ref{table:RAM_Elem}. There is good agreement between the DFT calculations, MD and experiment for the aluminum and copper potentials, where still the bulk modulus differs at most by 7\% between DFT and experiment. However, larger discrepancies are found between DFT and MD/experiment for silver and gold, sized to 15\% for the bulk modulus and 23\% for the cohesive energy. The variance between the DFT data and experiment closely follows known trends in DFT error and uncertainty quantification~\citep{lejaeghere_error_2014}. The original authors state that these potentials were fitted to lattice constants, elastic constants, bulk moduli, vacancy formation energies, and sublimation energies, but do not report the source of the fitting data. This comparison clearly implies that the elemental potentials are fitted to experimental data, and thus illustrates the need for scaling DFT training set data as described in Section~\ref{sec:dftset}. 

\begin{table}[htbp]
\begin{center}
\begin{tabular}{c c | c c c } 
Element & Property & DFT & MD & Exp.~\citep{kittel_introduction_2004} \\ [0.5ex] 
 \hline\hline
 & $E_{Coh}$ [eV] & 2.519 & 2.850 & 2.95 \\ 
 Ag & $a$ [\AA] & 4.145 & 4.093 & 4.09 \\
 & $K$ [GPa] & 90.581 & 102.880 & 100.7 \\
 \hline
 & $E_{Coh}$ [eV] & 3.548 & 3.580 & 3.39 \\
 Al & $a$ [\AA] & 4.041 & 4.053 & 4.05 \\
 & $K$ [GPa] & 77.712 & 72.590 & 72.2 \\
 \hline
 & $E_{Coh}$ [eV] & 3.035 & 3.930 & 3.81 \\
 Au & $a$ [\AA] & 4.156 & 4.084 & 4.08 \\
 & $K$ [GPa] & 139.315 & 165.640 & 173.2 \\
  \hline
 & $E_{Coh}$ [eV] & 3.485 & 3.540 & 3.49 \\
 Cu & $a$ [\AA] & 3.635 & 3.612 & 3.61 \\
 & $K$ [GPa] & 138.392 & 135.163 & 137 \\
\end{tabular}
\end{center}
\caption{Computed cohesive energy, lattice constant, and bulk modulus of elemental Ag, Al, Au, and Cu using DFT and MD with the standardized Zhou potentials used in RAMPAGE fitting.}
\label{table:RAM_Elem}
\end{table}

\begin{figure}[htbp]
         \centering
         \includegraphics[width=0.45\textwidth]{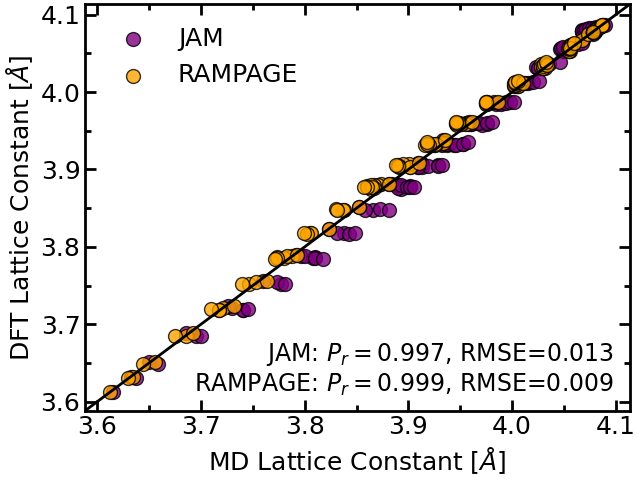}
         \label{fig:AgAuCu_QOI-lat}
     \hfill
         \centering
         \includegraphics[width=0.45\textwidth]{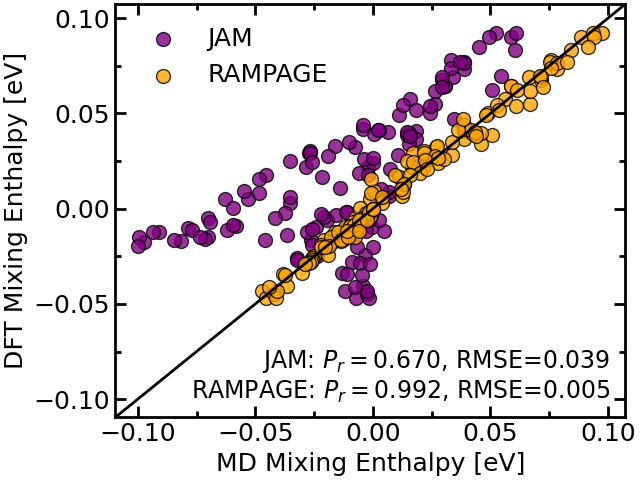}
         \label{fig:AgAuCu_QOI-mix}
     \vspace{1em}
         \centering
         \includegraphics[width=0.45\textwidth]{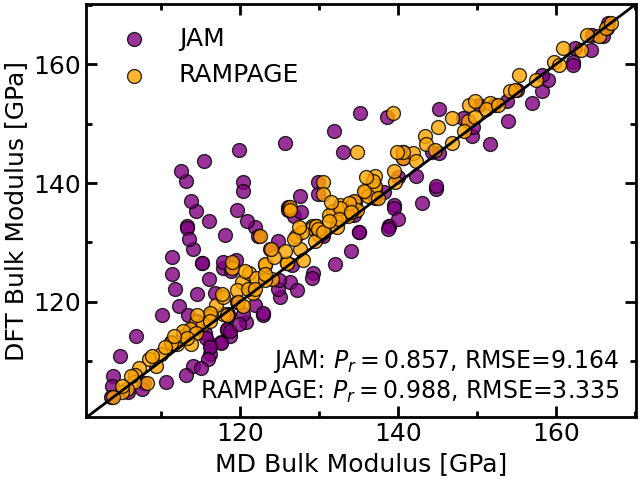}
         \label{fig:AgAuCu_QOI-bulk}
    \caption{Parity plots of lattice constant (top left), mixing enthlapy (top right), and bulk modulus (bottom) MD calculations versus DFT. Data is sampled across ternary composition space for Ag-Au-Cu RAMPAGE potentials (blue) and JAM (orange). The diagonal line corresponds to perfect matching between MD and DFT results.
    }
    \label{fig:AgAuCu_QOIparity}
\end{figure}

The combined ternary RAMPAGE Ag-Au-Cu potential is compared to JAM using equilibrium properties calculated across ternary composition space in the pairs plots in Figure~\ref{fig:AgAuCu_QOIparity}. The error weights and optimal training sets used for fitting the three required Ag-Au-Cu binaries is presented in the supplemental information. Across all three properties the RAMPAGE potential is able to closely reproduce the validation set DFT data. Recall that only binary properties are included in the training set, therefore these ternary structures may be considered part of the validation set. The lattice constant, which for solid solution typically follows linear behavior with some bowing, performs best for both RAMPAGE with a Pearson's correlation coefficient of 0.999 (JAM $P_r$=0.997). The mixing enthalpy and bulk modulus show similar trends. Mixing enthalpy, in particular is often tricky to accurately reproduce due to the quantity's small absolute value. As mixing enthalpy determines a system's tendency to form ordered intermetallics or phase separate, it is most important to ensure that the sign is accurately reproduced, which is seen in the RAMPAGE data.
RAMPAGE is able to accurately reproduce DFT data throughout composition space root mean square errors (RMSEs) uniformly lower than that produced by JAM potentials. The strong agreement between DFT and the RAMPAGE potential despite the deviation in elemental properties point to successful scaling of the training set data. 

\begin{figure}[htbp]
         \centering
         \includegraphics[width=0.48\textwidth]{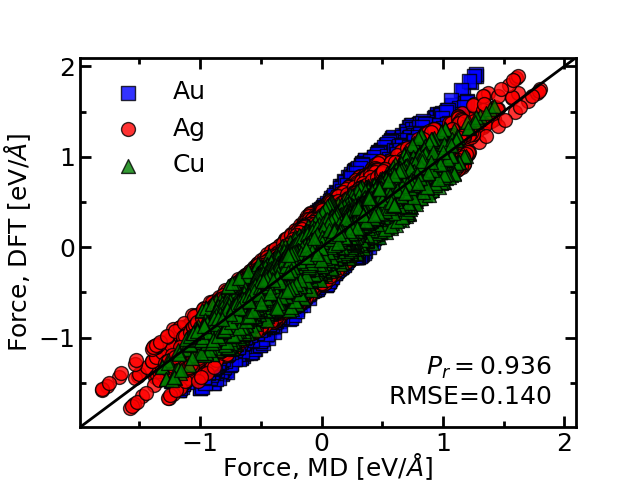}
         \label{fig:AgAuCu-300-RAMforce}
         \includegraphics[width=0.48\textwidth]{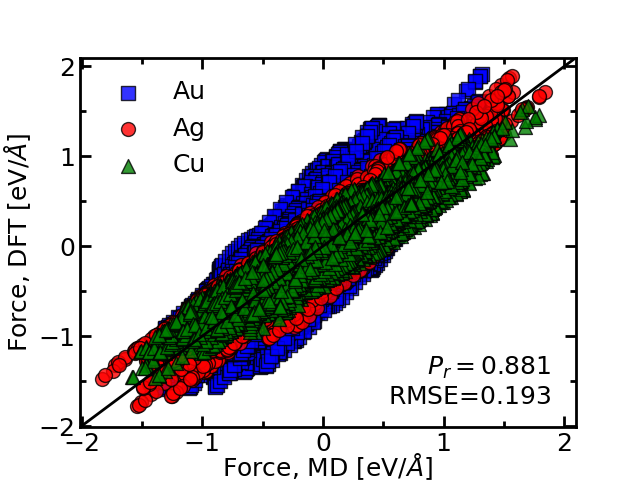}
         \label{fig:AgAuCu-300-JAMforce}
    \caption{Parity plots of forces computed with classical MD using (left) RAMPAGE potential and (right) JAM potential compared against DFT-MD. 108 atom cells were thermalized at 300~K. Data is shown for 500 time steps.}
    \label{fig:AgAuCu_forces}
\end{figure}

RAMPAGE potentials are fitted with 3 quantities: lattice parameter, which determines minimum energy atomic separation, the mixing enthalpy, which informs the depth of that energy well, and bulk modulus, which is dependant on the second derivative of the bottom of the energy well. The first derivative of the energy well, which corresponds to force is used in this work as a validation set. Force sets were generated by performing DFT-MD on an equiatomic 108~atom FCC SQSs thermalized to 300~K using an NVT-ensemble. The parity plots in Figure~\ref{fig:AgAuCu_forces} compare forces for RAMPAGE and JAM potentials against DFT. The RAMPAGE ternary is able to reproduce forces with a 27\% improvement in RMSE over JAM.   

Next we present results for the RAMPAGE potential when applied to the task of ripping atoms off (100) surface. The values of the computed energy barriers are presented in in Table~\ref{table:AgAuCu-ads}. The elemental potentials included sublimation energies in the training set; however, with the exception of gold, DFT more closely matches experiment~\citep{kittel_introduction_2004, tsong_field_1978}. The elemental sublimation energies have some of the largest deviation between MD and DFT ranging from 1.3~eV for silver to 0.2~eV for copper. 

\begin{table}[htbp]
\begin{center}
\begin{tabular}{c c c c c} 
 \hline
 & & \multicolumn{3}{c}{Energy [eV]} \\
 Adsorbate & Surface & MD & DFT & Exp.~\citep{tsong_field_1978} \\ 
 \hline\hline
  & Ag & -2.226 & -3.589 & 2.96 \\ 
 Ag & Au & -2.524 & -2.514 & --- \\
  & Cu & -2.476 & -2.543 & --- \\
  \hline
  & Ag & -3.165 & -2.931 & --- \\
 Au & Au & -3.535 & -3.050 & 3.78 \\ 
  & Cu & -3.861 & -3.354 & --- \\
  \hline
  & Ag & -2.511 & -2.551 & --- \\
 Cu & Au & -3.120 & -3.103 & --- \\
  & Cu & -2.751 & -2.976 & 3.50 \\ 
\end{tabular}
\end{center}
\caption{Energy barriers for ripping an atom from a 100 surface computed using RAMPAGE potentials and DFT-MD.}
\label{table:AgAuCu-ads}
\end{table}

Regardless of the surface, the largest deviations are observed from gold adsorbates, with a deviation of 0.2~eV for a silver surface and 0.5~eV for a copper surface. All other cross energy deviations are less than 0.1~eV. This would indicate that the selected gold elemental potential is relatively ill suited for simulations of sublimation or deposition, illustrating the need for careful selection of the constituent elemental potentials with respect to the target application. 
However, the cross interaction desorption energy deviations (with the exception of Au on Cu) are less than that of the elemental potentials
and the sublimation energy deviation for adsorbates not including gold are sufficiently low, thus we can conclude that the energies are well fitted within the limits of the elemental potential for this ternary case study.

\subsection{Ag-Al}
\label{sec:CaseStudyAgAl}

The Ag-Al system provides a significant challenge for simple interatomic potential models such as EAM due to its peculiar phase diagram. While there is an extended solubility of aluminium in silver, there is very limited solubility on the Al-rich end~\citep{lim_assessment_1995}. Nucleation and pre-nucleation in Al-rich Al-Ag alloys have attracted scientific interest for a significant time, and were one of the first systems in which Guinier found Guinier-Preston (GP) zones in 1942~\citep{guinier1942mecanisme}. Questions about these GP zones in Al-Ag such as their shape or faceting kept being discussed in the literature for many decades~\citep{alexander1984faceting}. Alloys with somewhat higher Ag concentrations also show peculiar dealloying behavior which can even be used to create nanostructured silver~\citep{wang2009influence}. These factors provide a difficult real-world test case to evaluate the performance of RAMPAGE potentials.

\begin{figure}[htbp]
    \centering
    \includegraphics[width=0.6\textwidth]{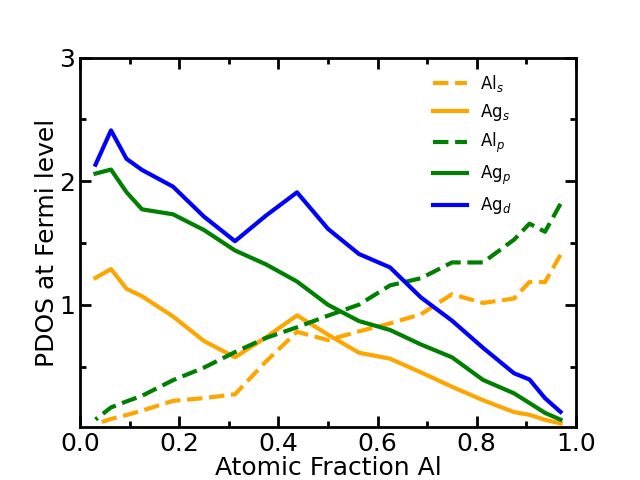}
    \caption{Partial density of states of Ag-Al 2x2x2 FCC SQS}
    \label{fig:PDOS}
\end{figure}

Since the lattice constants of Ag and Al are very similar to within 1\%, considering the Hume-Rothery (HR) rules, all these solubility effects should stem from the different valency (Al: +3, Ag: +1) and electronegativity (Ag: 1.93, Al: 1.61), and are thus chemical rather than geometrical effects. The different solubilities also follow the HR rule that a metal is more likely to dissolve a metal of higher valency, than vice versa. This can be quantified using \textit{ab initio} calculations. Figure~\ref{fig:PDOS} shows the partial density of states (PDOS) at the Fermi level of the 17 cell validation set. Below 65\% aluminum, bonding is dominated by the Ag-\textit{d} orbital, and above that point the bonding environment is primarily dominated by \textit{p}-shell electrons. 

It was observed in the process of fitting potentials for the Ag-Al system that producing a single binary potential to describe both sides of composition space results in unacceptable compromises in the resulting QOIs. A simple model such as EAM-FS is unable to account for the change in bonding environment from \textit{d}- to \textit{p}-shell electrons. Thus, two RAMPAGE potentials were generated, each optimized to Ag- and Al-rich alloys. The training sets consisted of SQSs of 0.03125, 0.125 0.4375, 0.5625, 0.75, 0.88, and 0.96875 atomic fraction Al. A logistic function using the atomic fractions as input and set with an inflection point at 0.5 was used to generate error weights $w_i : (0,1)$ in Eqn.\ ~\ref{eq:fitnessFunction}. 
With the objective functions biased towards either Al-rich or Ag-rich alloys, the fitting and selection procedure remains largely unchanged from that discussed in Sec.\  \ref{ssec:FitPrtoc} and the supplemental information. We find that the best fits was achieved by excluding intermetallic structures from the training set. 

\begin{figure}[htbp]
    \centering
         \centering
         \includegraphics[width=0.49\textwidth]{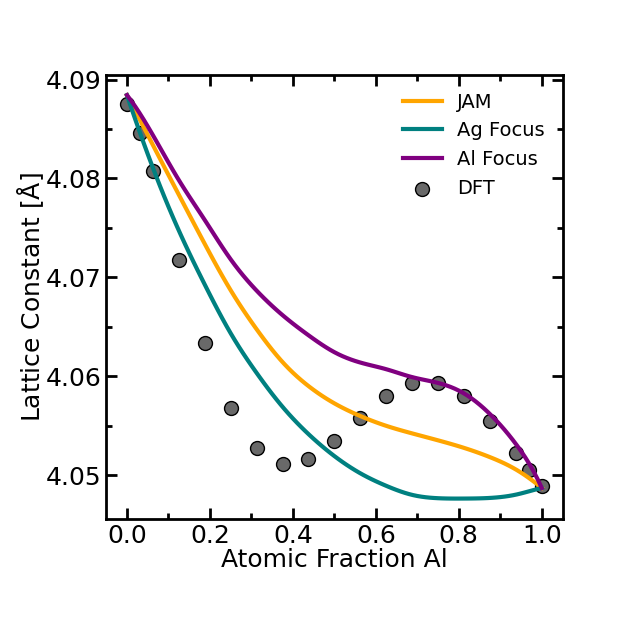}
         \label{fig:AgAlLat}
         \centering
         \includegraphics[width=0.49\textwidth]{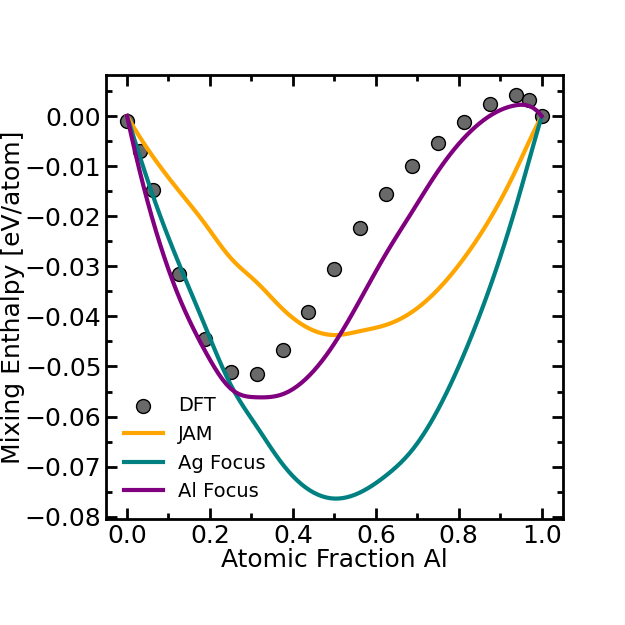}
         \label{fig:AgAlMix}
         \centering
         \includegraphics[width=0.49\textwidth]{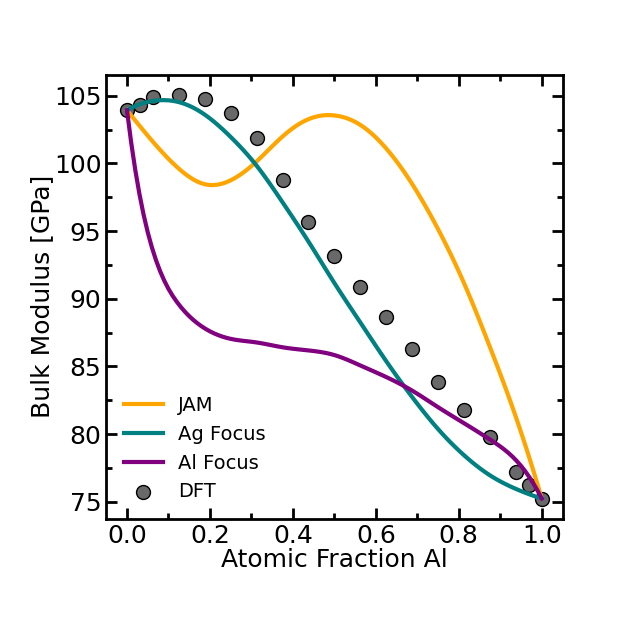}
         \label{fig:AgAlBulk}
    \caption{Lattice constant, mixing enthalpy, and bulk modulus for Ag-Al alloys computed using DFT, MD with RAMPAGE a fitted potential, and MD with a JAM potential.}
    \label{fig:AgAlQOI}
\end{figure}

The optimum Ag-focued fit was produced with an error weight set of $w_H=1.6$, $w_K=1.2$, and $w_a=1.6$ which biases the objective function away from bulk modulus, whereas the error weights that produced the best Al-focued fit put the greatest emphasis on lattice constant and least on bulk modulus with $w_H=1.0$, $w_K=0.8$, and $w_a=1.2$. The QOIs calculated by DFT, JAM, and the two RAMPAGE fits are shown in Figure \ref{fig:AgAlQOI}. Each fit is designed to reproduce the QOIs to within 0.2 atomic fraction solubility. When considering only these regions, the RAMPAGE fits produce RMSEs an order of magnitude lower (0.37 and 0.31) than JAM (2.55 and 4.13) for Ag- and Al-rich alloys respectively. Both fits are well optimized and reproduce the validation set within their intended composition ranges. 
 
\begin{figure}[htbp]
          \centering
         \includegraphics[width=0.49\textwidth]{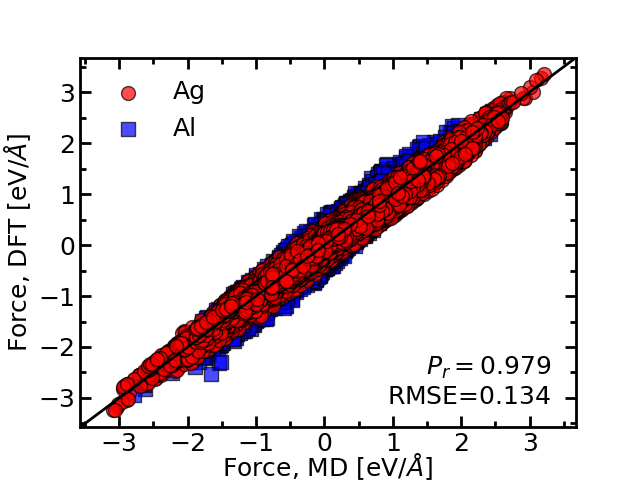}
         \label{fig:AgAlForcesRAM}
         \includegraphics[width=0.49\textwidth]{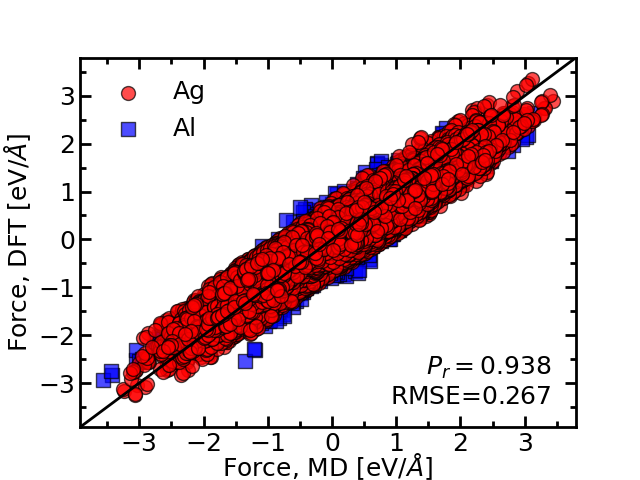}
         \label{fig:AgAlForcesJAM}        
         
         \centering
         \includegraphics[width=0.49\textwidth]{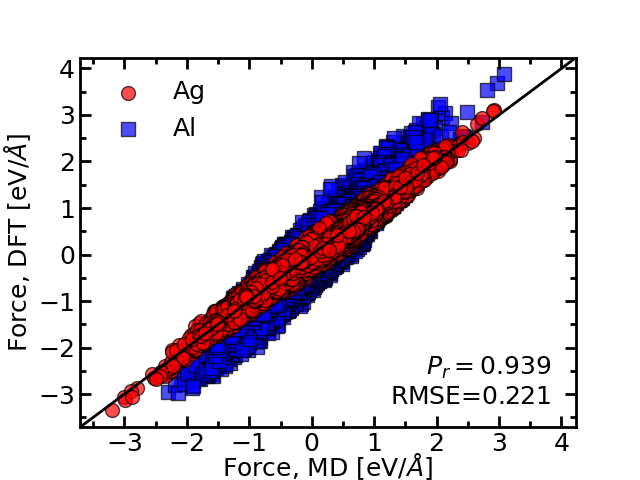}
         \includegraphics[width=0.49\textwidth]{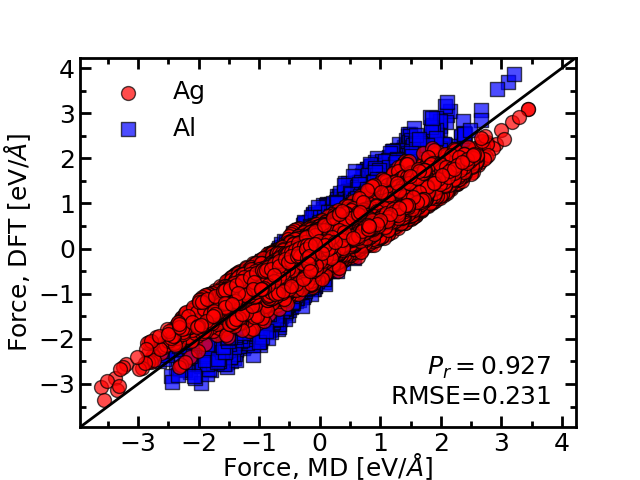}
    \caption{Parity plots of computed forces using (left) RAMPAGE Al-focus and (right) JAM potentials against DFT for solid solutions (top) Ag$_{81}$Al$_{27}$ and (bottom) Ag$_{24}$Al$_{84}$ at 840~K.}
    \label{fig:AgAlForces}
\end{figure}

To appropriately assess the forces resulting from the fitted cross interactions, and minimize contributions from the elemental potentials, forces are computed at the limits of solubility: using 108-atom $Ag_{75}Al_{25}$ and $Ag_{22}Al_{78}$ 
cells at 840~K. The elemental silver and aluminum potentials are well behaved with respect to DFT computed forces producing correlation coefficients of 0.997 and 0.990 and RMSEs of 0.031 and 0.091 respectively. These can be taken as the upper limits for force accuracy for the potentials in this case study. The pairs plots in Figure~\ref{fig:AgAlForces} show that the Al-focused RAMPAGE fit is well calibrated with respect to forces when compared to JAM. The Ag-focused fit produces a correlation coefficient of 0.938 and RMSE of 0.352 for $Ag_{75}Al_{25}$. The RMSE is a factor of two larger than the Al-focused fit. While while the Al-focused fit reproduces forces of Ag-rich alloys, the large deviation in bulk modulus at high Ag concentrations makes it ill-suited in that concentration rage. The intended target of this case study is simulating atomic-probe tomography simulations, and the process of ripping atoms from a surface is dependant on the elastic response of the material. 

To benchmark the quality of the fitted potential for use in APT simulations, the desorption energies are computed and sown in Table~\ref{table:AgAlAds}. As discussed in the previous section, the largest discrepancy between the RAMPAGE potential and DFT comes from the elemental interactions in removing a silver atom from a silver surface. Each RAMPAGE potential is capable of describing the bonding environment of its associated training set with a deviation of 5.6\% for an Ag adsorbate on an Al surface using the Al-focused fit and 0.2\% for an Al adsorbate on an Ag surface with the Ag-focused fit. The RAMPAGE fits also show acceptable agreement with DFT when used in the bonding environment not included in the training set, albeit with understandably higher errors (16\% for the Ag-focused fit and 12\% for the Al-focused fit). 

\begin{table}[htbp]
\begin{center}
\begin{tabular}{c c c c c c} 
\hline
 & & \multicolumn{4}{c}{Energy [eV]} \\
 Adsorbate & Surface & Ag-focus & Al-Focus & DFT & Exp.~\citep{tsong_field_1978} \\ 
\hline\hline
\multirow{2}*{Ag} & Ag & -2.226 & -2.226 & -3.589 & -2.96 \\ 
  & Al & -2.437 & -2.893 & -2.899 & --- \\
\hline 
 \multirow{2}*{Al} & Ag & -3.148 & -3.750 &-3.337 & --- \\
  & Al & -3.049 & -3.049 & -3.099 & -3.34 \\ 
\end{tabular}
\end{center}
\caption{Energy for sublimating an adsorbate from a 100 surface. Computed using RAMPAGE potentials and DFT-MD. The Ag-Ag data is unchanged from Table~\ref{table:AgAuCu-ads}.}
\label{table:AgAlAds}
\end{table}

The reproduction of this validation set data leads us to conclude that RAMPAGE fit is able to successfully model Al-rich Ag-Al alloys. A truly qualitative data-driven understanding of the model error and quality of the resulting interatomic potential requires a far more advanced statistical framework than that presented here. The complexity associated with a rigorous uncertainty quantification put it outside the scope of this work. A Bayesian framework for evaluating uncertainty in interatomic potentials can be found in ref.~\citep{hedge_bayesian_2022}. However, within the this property driven framework, the validation evidence points to a well calibrated model.

\section{Conclusions}
In this work, we present and describe the RAMPAGE code package for fitting multi-component interatomic potentials from published elemental potentials. Within the RAMPAGE framework, off-the-shelf elemental potentials are used as model inputs to efficiently fit cross interactions for multi-component alloy potentials. Binary Ag-Al and ternary Ag-Au-Cu case studies are used to present and benchmark the resulting fitted potentials. RAMPAGE potentials are fitted and compared against JAM potentials to show the performance of RAMPAGE potentials in predicting equilibrium properties and forces. It is demonstrated that for both well behaved, and immiscible alloy systems, RAMPAGE potentials are capable of reproducing equilibrium properties and forces. Due to the differing valances of silver and aluminum, two potentials are required to sufficiently model each solid-solution region. It was shown that RAMPAGE was able to bias the objective function to fit composition dependant potentials with a general training set. 


We then apply the RAMPAGE potentials to the problem of atomic evaporation on a (100) surface to benchmark the utility of RAMPAGE potentials in simulating APT and PVD processes. It was demonstrated that RAMPAGE potentials are able to model such processes with accuracy that exceeds that of the elemental potential inputs. Thus, RAMPAGE potentials are able to successfully capture cross interaction behavior for alloy systems with simulations not included in the training set of either the binary or elemental potentials. Therefore we conclude that the RAMPAGE fitting procedure is able to produce functional high efficiency EAM interatomic potentials and enable the modeling of multi-component systems which currently suffer from a lack of suitable potentials. 

\section{Acknowledgements}

This work was funded by the Air Force Office of Scientific Research (AFOSR) under Grant No.~FA9550-19-1-0378. Simulations were performed at the Ohio Supercomputer Center (Grant No PAS0072).


 \bibliographystyle{unsrt} 
 \bibliography{references}

\begin{thebibliography}{10}

\bibitem{sikandar_mashayak_lammps_nodate}
{Sikandar Mashayak}, {Mike Brown}, {Carl Ponder}, {Christian Trott}, {Paul
  Crozier}, {Fiona Reid}, and {Courtenay Vaughan}.
\newblock {LAMMPS} {Benchmarks}, 2016.

\bibitem{becker_considerations_2013}
Chandler~A. Becker, Francesca Tavazza, Zachary~T. Trautt, and Robert~A.
  Buarque~de Macedo.
\newblock Considerations for choosing and using force fields and interatomic
  potentials in materials science and engineering.
\newblock {\em Current Opinion in Solid State and Materials Science},
  17(6):277--283, December 2013.

\bibitem{hale_evaluating_2018}
Lucas~M. Hale, Zachary~T. Trautt, and Chandler~A. Becker.
\newblock Evaluating variability with atomistic simulations: the effect of
  potential and calculation methodology on the modeling of lattice and elastic
  constants.
\newblock {\em Modelling and Simulation in Materials Science and Engineering},
  26(5):055003, May 2018.
\newblock Publisher: IOP Publishing.

\bibitem{tadmor2011potential}
Ellad~B Tadmor, Ryan~S Elliott, James~P Sethna, Ronald~E Miller, and Chandler~A
  Becker.
\newblock The potential of atomistic simulations and the knowledgebase of
  interatomic models.
\newblock {\em JOM}, 63(7):17, 2011.

\bibitem{gardner_use_2005}
Leroy Gardner.
\newblock The use of stainless steel in structures.
\newblock {\em Progress in Structural Engineering and Materials}, 7(2):45--55,
  2005.
\newblock \_eprint: https://onlinelibrary.wiley.com/doi/pdf/10.1002/pse.190.

\bibitem{ryan_why_2002}
Mary~P. Ryan, David~E. Williams, Richard~J. Chater, Bernie~M. Hutton, and
  David~S. McPhail.
\newblock Why stainless steel corrodes.
\newblock {\em Nature}, 415(6873):770--774, February 2002.
\newblock Number: 6873 Publisher: Nature Publishing Group.

\bibitem{thompson2015spectral}
Aidan~P Thompson, Laura~P Swiler, Christian~R Trott, Stephen~M Foiles, and
  Garritt~J Tucker.
\newblock Spectral neighbor analysis method for automated generation of
  quantum-accurate interatomic potentials.
\newblock {\em Journal of Computational Physics}, 285:316--330, 2015.

\bibitem{chenoweth_reaxff_2008}
Kimberly Chenoweth, Adri C.~T. van Duin, and William~A. Goddard.
\newblock {ReaxFF} {Reactive} {Force} {Field} for {Molecular} {Dynamics}
  {Simulations} of {Hydrocarbon} {Oxidation}.
\newblock {\em The Journal of Physical Chemistry A}, 112(5):1040--1053,
  February 2008.
\newblock Publisher: American Chemical Society.

\bibitem{bartok2010gaussian}
Albert~P Bart{\'o}k, Mike~C Payne, Risi Kondor, and G{\'a}bor Cs{\'a}nyi.
\newblock Gaussian approximation potentials: The accuracy of quantum mechanics,
  without the electrons.
\newblock {\em Physical review letters}, 104(13):136403, 2010.

\bibitem{plimpton_computational_2012}
Steven~J. Plimpton and Aidan~P. Thompson.
\newblock Computational aspects of many-body potentials.
\newblock {\em MRS Bulletin}, 37(5):513--521, May 2012.
\newblock Publisher: Cambridge University Press.

\bibitem{daw_embedded-atom_1984}
Murray~S. Daw and M.~I. Baskes.
\newblock Embedded-atom method: {Derivation} and application to impurities,
  surfaces, and other defects in metals.
\newblock {\em Physical Review B}, 29(12):6443--6453, June 1984.

\bibitem{zhou_misfit-energy-increasing_2004}
X.~W. Zhou, R.~A. Johnson, and H.~N.~G. Wadley.
\newblock Misfit-energy-increasing dislocations in vapor-deposited
  {CoFe}/{NiFe} multilayers.
\newblock {\em Physical Review B}, 69(14):144113, April 2004.

\bibitem{sheng_highly_2011}
H.~W. Sheng, M.~J. Kramer, A.~Cadien, T.~Fujita, and M.~W. Chen.
\newblock Highly optimized embedded-atom-method potentials for fourteen fcc
  metals.
\newblock {\em Physical Review B}, 83(13):134118, April 2011.

\bibitem{johnson_alloy_1989}
R.~A. Johnson.
\newblock Alloy models with the embedded-atom method.
\newblock {\em Physical Review B}, 39(17):12554--12559, June 1989.

\bibitem{parviainen_atomistic_2015}
S.~Parviainen, F.~Djurabekova, S.~P. Fitzgerald, A.~Ruzibaev, and K.~Nordlund.
\newblock Atomistic simulations of field assisted evaporation in atom probe
  tomography.
\newblock {\em Journal of Physics D: Applied Physics}, 49(4):045302, December
  2015.
\newblock Publisher: IOP Publishing.

\bibitem{yao_effects_2015}
Lan Yao, Travis Withrow, Oscar~D. Restrepo, Wolfgang Windl, and Emmanuelle~A.
  Marquis.
\newblock Effects of the local structure dependence of evaporation fields on
  field evaporation behavior.
\newblock {\em Applied Physics Letters}, 107(24):241602, December 2015.
\newblock Publisher: American Institute of Physics.

\bibitem{oberdorfer_influence_2018}
C.~Oberdorfer, T.~Withrow, L.~J. Yu, K.~Fisher, E.~A. Marquis, and W.~Windl.
\newblock Influence of surface relaxation on solute atoms positioning within
  atom probe tomography reconstructions.
\newblock {\em Materials Characterization}, 146:324--335, December 2018.

\bibitem{qi_ab-initio_2022}
Jiayuwen Qi, Christian Oberdorfer, Emmanuelle~A. Marquis, and Wolfgang Windl.
\newblock Ab-{Initio} {Simulation} of {Field} {Evaporation}, July 2022.
\newblock arXiv:2207.03958 [cond-mat].

\bibitem{daw_semiempirical_1983}
Murray~S. Daw and M.~I. Baskes.
\newblock Semiempirical, {Quantum} {Mechanical} {Calculation} of {Hydrogen}
  {Embrittlement} in {Metals}.
\newblock {\em Physical Review Letters}, 50(17):1285--1288, April 1983.

\bibitem{finnis_simple_1984}
M.~W. Finnis and J.~E. Sinclair.
\newblock A simple empirical {N}-body potential for transition metals.
\newblock {\em Philosophical Magazine A}, 50(1):45--55, July 1984.

\bibitem{wadley_mechanisms_2001}
H.~N.~G Wadley, X~Zhou, R.~A Johnson, and M~Neurock.
\newblock Mechanisms, models and methods of vapor deposition.
\newblock {\em Progress in Materials Science}, 46(3):329--377, January 2001.

\bibitem{voter1994embedded}
Arthur~F Voter.
\newblock The embedded atom method.
\newblock {\em Intermetallic Compounds: Principles}, 1:77, 1994.

\bibitem{ercolessi1986glue}
F~Ercolessi, E~Tosatti, and M~Parrinello.
\newblock Au (100) surface reconstruction.
\newblock {\em Physical review letters}, 57(6):719, 1986.

\bibitem{pun_development_2009}
G.~P.~Purja Pun and Y.~Mishin.
\newblock Development of an interatomic potential for the {Ni}-{Al} system.
\newblock {\em Philosophical Magazine}, 89(34-36):3245--3267, December 2009.

\bibitem{williams_embedded-atom_2006}
P.~L. Williams, Y.~Mishin, and J.~C. Hamilton.
\newblock An embedded-atom potential for the {Cu}–{Ag} system.
\newblock {\em Modelling and Simulation in Materials Science and Engineering},
  14(5):817, 2006.

\bibitem{mendelev_development_2009}
M.I. Mendelev, M.J. Kramer, R.T. Ott, D.J. Sordelet, D.~Yagodin, and P.~Popel.
\newblock Development of suitable interatomic potentials for simulation of
  liquid and amorphous {Cu}–{Zr} alloys.
\newblock {\em Philosophical Magazine}, 89(11):967--987, April 2009.
\newblock Publisher: Taylor \& Francis \_eprint:
  https://doi.org/10.1080/14786430902832773.

\bibitem{cheng_atomic_2009}
Y.~Q. Cheng, E.~Ma, and H.~W. Sheng.
\newblock Atomic {Level} {Structure} in {Multicomponent} {Bulk} {Metallic}
  {Glass}.
\newblock {\em Physical Review Letters}, 102(24):245501, June 2009.

\bibitem{landa_development_1998}
A.~Landa, P.~Wynblatt, A.~Girshick, V.~Vitek, A.~Ruban, and H.~Skriver.
\newblock Development of {Finnis}–{Sinclair} type potentials for {Pb},
  {Pb}–{Bi}, and {Pb}–{Ni} systems: application to surface segregation.
\newblock {\em Acta Materialia}, 46(9):3027--3032, May 1998.

\bibitem{koleske_molecular_1993}
D.~D. Koleske and S.~J. Sibener.
\newblock Molecular dynamics simulations of the basal planes of {Ni} and {Cu}
  using {Finnis}-{Sinclair} potentials.
\newblock {\em Surface Science}, 290(1):179--194, June 1993.

\bibitem{hyodo_empirical_2020}
Katsutoshi Hyodo, Shinji Munetoh, Toshihiro Tsuchiyama, and Setsuo Takaki.
\newblock Empirical interatomic potential for {Fe}-{N} binary system based on
  {Finnis}–{Sinclair} potential.
\newblock {\em Computational Materials Science}, 174:109500, March 2020.

\bibitem{voter_accurate_1986}
Arthur~F. Voter and Shao~Ping Chen.
\newblock Accurate {Interatomic} {Potentials} for {Ni}, {Al} and {Ni3Al}.
\newblock {\em MRS Online Proceedings Library Archive}, 82, 1986.

\bibitem{voter1993embedded}
Arthur~F Voter.
\newblock Embedded atom method potentials for seven fcc metals: Ni, pd, pt, cu,
  ag, au, and al.
\newblock {\em Los Alamos Unclassified Technical Report\# LA-UR}, pages
  93--3901, 1993.

\bibitem{chantasiriwan_higher-order_1996}
Somchart Chantasiriwan.
\newblock Higher-order elasticity of cubic metals in the embedded-atom method.
\newblock {\em Physical Review B}, 53(21):14080--14088, 1996.

\bibitem{kresse_ab_1993}
G.~Kresse and J.~Hafner.
\newblock Ab initio molecular dynamics for liquid metals.
\newblock {\em Physical Review B}, 47(1):558--561, January 1993.
\newblock Publisher: American Physical Society.

\bibitem{kresse_ab_1994}
G.~Kresse and J.~Hafner.
\newblock Ab initio molecular-dynamics simulation of the
  liquid-metal--amorphous-semiconductor transition in germanium.
\newblock {\em Physical Review B}, 49(20):14251--14269, May 1994.
\newblock Publisher: American Physical Society.

\bibitem{blochl_projector_1994}
P.~E. Blöchl.
\newblock Projector augmented-wave method.
\newblock {\em Physical Review B}, 50(24):17953--17979, December 1994.
\newblock Publisher: American Physical Society.

\bibitem{perdew_generalized_1996}
John~P. Perdew, Kieron Burke, and Matthias Ernzerhof.
\newblock Generalized {Gradient} {Approximation} {Made} {Simple}.
\newblock {\em Physical Review Letters}, 77(18):3865--3868, October 1996.
\newblock Publisher: American Physical Society.

\bibitem{perdew_generalized_1997}
John~P. Perdew, Kieron Burke, and Matthias Ernzerhof.
\newblock Generalized {Gradient} {Approximation} {Made} {Simple}.
\newblock {\em Physical Review Letters}, 78(7):1396--1396, February 1997.

\bibitem{plimpton_fast_1995}
Steve Plimpton.
\newblock Fast {Parallel} {Algorithms} for {Short}-{Range} {Molecular}
  {Dynamics}.
\newblock {\em Journal of Computational Physics}, 117(1):1--19, March 1995.

\bibitem{hedge_bayesian_2022}
Arun Hegde, Cosmin Safta, Habib Najm, Elan Weiss, and Wolfgang Windl.
\newblock Bayesian calibration of interatomic potential models for binary
  alloys.
\newblock Accepted and pending publication in Computational Materials Science,
  2022.

\bibitem{asm_phase_2016}
H.~Okamoto, M.E. Schlesinger, and E.M. Mueller.
\newblock {\em {Alloy Phase Diagrams}}.
\newblock ASM International, 04 2016.

\bibitem{lejaeghere_error_2014}
K.~Lejaeghere, V.~Van~Speybroeck, G.~Van~Oost, and S.~Cottenier.
\newblock Error {Estimates} for {Solid}-{State} {Density}-{Functional} {Theory}
  {Predictions}: {An} {Overview} by {Means} of the {Ground}-{State} {Elemental}
  {Crystals}.
\newblock {\em Critical Reviews in Solid State and Materials Sciences},
  39(1):1--24, January 2014.
\newblock Publisher: Taylor \& Francis \_eprint:
  https://doi.org/10.1080/10408436.2013.772503.

\bibitem{kittel_introduction_2004}
Charles Kittel.
\newblock {\em Introduction to {Solid} {State} {Physics}}.
\newblock Wiley, Hoboken, NJ, 8 edition edition, November 2004.

\bibitem{tsong_field_1978}
T.~T. Tsong.
\newblock Field ion image formation.
\newblock {\em Surface Science}, 70(1):211--233, January 1978.

\bibitem{lim_assessment_1995}
S.~S. Lim, P.~L. Rossiter, and J.~E. Tibballs.
\newblock Assessment of the {Al}-{Ag} binary phase diagram.
\newblock {\em Calphad}, 19(2):131--141, June 1995.

\bibitem{guinier1942mecanisme}
Andr{\'e} Guinier.
\newblock Le m{\'e}canisme de la pr{\'e}cipitation dans un cristal de solution
  solide m{\'e}tallique.-cas des syst{\`e}mes aluminium-cuivre et
  aluminium-argent.
\newblock {\em J. Phys. Radium}, 3(7):124--136, 1942.

\bibitem{alexander1984faceting}
KB~Alexander, FK~LeGoues, HI~Aaronson, and DE~Laughlin.
\newblock Faceting of gp zones in an al-ag alloy.
\newblock {\em Acta Metallurgica}, 32(12):2241--2249, 1984.

\bibitem{wang2009influence}
Xiaoguang Wang, Zhen Qi, Changchun Zhao, Weimin Wang, and Zhonghua Zhang.
\newblock Influence of alloy composition and dealloying solution on the
  formation and microstructure of monolithic nanoporous silver through chemical
  dealloying of al- ag alloys.
\newblock {\em The Journal of Physical Chemistry C}, 113(30):13139--13150,
  2009.

\end{thebibliography}

\end{document}


\section{RAMPAGE Input File Format}
\label{sec:RAMPAGEinput}

The RAMPAGE fitting process for multicomponent EAM potentials can be broken down into three steps. First, elemental potentials are standardized, transformed, and then combined into baseline binary potentials. Second, the RAMPAGE fitting engine is employed to fit cross interactions to equilibrium properties. Third, the fitted binaries are assembled into multicompoent potentials. For example, to fit a four component A-B-C-D potential, the four chosen elemental potentials must be standardized, then binary systems A-B, A-C, A-D, B-C, B-D, and C-D must be fitted and combined. The pre and post processing steps are handled by scripts, and the fitting is handled by a compiled fitting engine that takes in an input file. What follows is a sample input file to the RAMPAGE fitting engine for a hypothetical A-B Finnis-Sinclair system. 

\begin{Verbatim} [fontsize=\small]
27 # Number of parallel processes
Finnis # EAM model to use: [Finnis] or [Alloy]
Morse # Pair potential function
Interpolate # Electron density function (line ignored for EAM Alloy)
2 # Number of components
A -3.00 # [Element] and [Energy of atom in equilibrium structure]
B -6.00
8 # Number of training structures
# [Name] [#Atoms] [#A] [#B] [#Cells] [Error Weight]
B2    2   1  1 343 1.0 
L1_2  4   3  1 125 1.0
L1_2i 4   1  3 125 1.0
sqs03 32 31  1  27 1.0
sqs25 32 24  8  27 1.0
sqs50 32 16 16  27 1.0
sqs75 32  8 24  27 1.0
sqs97 32  1 31  27 1.0
# Data for each structure [Lat. Constant] [H mix] [Bulk Mod.]
#B2# 
#L1_2#
#L1_2i#
#sqs03# 
#sqs25#
#sqs50#
#sqs75#
#sqs97#
# [guess] [max step] [lower bound] [upper bound]
5.0e-1 2.0e-1 0.0e+0 1.0e+0 # S_a 
5.0e-1 2.0e-1 0.0e+0 1.0e+0 # S_b
2.5e-1 1.0e-1 0.0e+0 1.0e+1 # D_e
1.5e+0 6.8e-1 1.0e-1 1.0e+2 # alpha
2.9e+0 1.3e+0 0.0e+0 1.0e+2 # r_e
1.0 # weight factor for mixing enthalpy
1.0 # weight factor for bulk modulus
1.0 # weight factor for lattice constant
101  # Number of batches
75 # Number of samples per batch
EGenetic # Algorithm
1000 # Initial pool size
50 # Pool size
40 # Number Top
0.90 # Mutation Prob Start
0.003 # Mutation Prob Decay (Linear)
0.50 # Mutation Prob Minimum
4    # FOM Exaggeration Factor
9   # Number of pressure states for bulk modulus (values given in next line)
20000 15000 10000 5000 0 -5000 -10000 -15000 -20000
\end{Verbatim}

The first line of the input file is the number of secondary processes to be run in parallel, and may take on a maximum value of one minus the number of processors available. At least one processor must be held in reserve for main RAMAPGE instance. The following three lines define the EAM model to be used and the functional form of the cross interactions. The next lines define the number of elements in the fit, followed by a line for each element defining the element symbol and the cohesive energy of the equilibrium structured given by the elemental potential in eV. 

The next block defines the number of training structures as well as a line containing information on each structure. Immediately after the structure information is a block for the training set data. In the example above, placeholders are used. When running RAMPAGE, these placeholders must be replaced by the lattice constant in Angstroms, mixing enthalpy in eV, and bulk modulus in GPa. These three numbers are separated by white-space. The structures are defined in a `./TEMPLATE/' directory, which must be located in the current working directory. Standard LAMMPS input structure is expected with a naming convention of `data.template\_\#NAME\#' where `\#NAME\#' is the structure label to be used in the input file. 

The initial guess, max step size, and bounds for the fitting parameters follow the training set data block. The number of fitting parameters and their order are defined as part of the selected pair potential and electron density functional forms. These are listed in \ref{sec:functions}. The three error weight factors act on Equation~12. Their selection is non-trivial and discussed in \ref{sec:errorWeights}. The remaining lines in the input file define options for the fitting algorithm and typically do not need to be changed.

\section{Pair Potential and Electron Density Functions}
\label{sec:functions}

Within RAMPAGE, one of several functional forms for the pair potential and electron density may be selected on lines three and four of the input file. Here we list the functions currently implemented within RAMPAGE and their parameters.

\subsection{Pair Potential Functions}

Pair potential functions are selected with the 3rd line of the RAMPAGE input file. The following are the available options.

``Morse" selects a standard 3 parameter Morse function, which is shown in Equation~6 in the main text and is repeated below. The input file expects the parameters to be in the following order: $D_e$, $\alpha$, $r_e$.
\begin{equation}
\label{eq:morseSM}
V_{AB}(r) = D_e \left( \exp \left( -2 \alpha \left( r - r_{e} \right) \right) - 2 \exp \left( - \alpha \left( r - r_{e} \right) \right) \right).
\end{equation}
``GenMorse" selects a more generalized form of the Morse function with 5 fitting parameters. The general Morse function reduces to the standard Morse function when $\beta = 2$ and $\delta = 0$. The ordering of parameters in the input file is: $D_e$, $\alpha$, $\beta$ ,$r_e$, $\delta$.
\begin{equation}
\label{eq:GenMorseSM}
V_{AB}(r) = D_e \left( \exp \left( -\beta \alpha \left( r - r_{e} \right) \right) - \beta \exp \left( - \alpha \left( r - r_{e} \right) \right) \right) + \delta
\end{equation}
``RosenMorse" selects a 3 parameter Rosen-Morse function, where the parameters in order are: $V_1$, $V_2$, and $\alpha$
\begin{equation}
\label{eq:rosenmorseSM}
V_{AB}(r) = V_{1} \tanh \left( \frac{r}{\alpha} \right) - V_{1} \sech ^{2} \left( \frac{r}{\alpha} \right)
\end{equation}
``ModRMorse" selects a 5 parameter modified Rosen-Morse function which takes the following form:
\begin{equation}
\label{eq:modrosenmorseSM}
V_{AB}(r) = D_e \left ( 1 - \frac{\exp(\alpha (r_e - r_{ij})) + 1}{\exp(\alpha (r - r_{ij})) + 1} \right ),
\end{equation}
where $r_{ij} = r_{e} - \sqrt{(KD_e)/k_e}$. $K$ is a dimensionless constant, and $k_e$ is the equilibrium harmonic vibrational force constant between two species which may be treated as a free parameter or explicitly computed and fixed by limiting the upper and lower bounds. The parameter ordering is: $k_e$, $K$, $D_e$, $\alpha$, $r_e$
``DCIModRMorse" selects the 3 parameter Doug Cruz-Irrison modified Rosen-Morse potential with parameters $V_1$, $V_2$, and $a$. The form of the potential is,
\begin{equation}
\label{eq:dci_modrosenmorseSM}
V_{AB}(r) = - \frac{V_1 - V_2 \sinh(r/\alpha)}{\cosh^2(r/\alpha)}.
\end{equation}

\subsection{Electron Density Functions}

The form of the electron density function is chosen in line 4 of the input file. If fitting an EAM/alloy potential, this line is ignored. For EAM/FS potentials, parameters denoted with the subscript `A' determine $\rho_{AB}$ parameters with subscript `B' fit $\rho_{BA}$. The parameters of the electron density function always preceded those of the pair potential in the input file when choosing initial starting guesses, max step size, and bounds for fitting. 

The keyword ``Scaled'' selects an electron density form where cross interactions are given by scaling the elemental electron densities. This model contains two free parameters: $S_A$ and $S_B$, and takes the following form
\begin{equation}
\label{eq:scaledSM}
\rho_{AB}(r) = S_A\rho_{AA}(r).
\end{equation}
``Interpolated" selects a 2 parameter model where cross-species electron densities are determined by linear interpolation/extrapolation of the elemental electron densities as shown in Equation \ref{eq:interpolatedSW}. When the free parameters $S_A$ and $S_B$ are between 0 and 1, the resulting electron density is a linear interpolation between the two elemental electron densities. Beyond that range, the electron densities are extrapolated.
\begin{equation}
\label{eq:interpolatedSW}
\rho_{AB}(r) = \rho_{AA}(r) + S_A(\rho_{AA}-\rho_{BB}),
\end{equation}
``Voter" selects an analytical form for the electron density based on the hydrogenic 4s orbital as suggested by Voter in 1986.  This model contains 2 free parameters, $a_A$ and $a_B$, and takes the following form
\begin{equation}
\label{eq:voterSW}
\rho_{AB}(r) = r^6 \left( \exp(-a_A r) + 2^9\exp(-2 a_A r) \right).
\end{equation}
``GenVoter" is a modified more general form of the Voter electron density with an additional scaling parameter $S$. The parameters for this model take the order $S_A$, $S_B$, $a_A$, $a_B$ in the RAMPAGE input file.
\begin{equation}
\label{eq:genVoterSW}
\rho_{AB}(r) = S_A r^6 \left( \exp(-a_A r) + 2^9\exp(-2 a_A r) \right)
\end{equation}
``GenVoter2" takes a form similar to that above, but now has free parameters $S_A$, $S_B$, $a_A$, $a_B$, $b_A$, $b_B$
\begin{equation}
\label{eq:genVoter2SW}
\rho_{AB}(r) = S_A r^6 \left( \exp(-a_A r) + 2^9\exp(-b_A r) \right)
\end{equation}
``CSW" selects an 4 parameter function from Chantasiriwan and Milstein, with fitting parameters $a$, $a_1$, $a_2$, $b$:
\begin{equation}
\label{eq:csw_appx}
\rho_{AB}(r) = \frac{1+a_1 \cos (a_0 r) + a_{2} \sin (a_0 r)}{r^b},
\end{equation}

\section{Selecting Error Weights and Training Set}
\label{sec:errorWeights}

RAMPAGE uses a weighted sum construction to balance the three training set properties, and the weight factors $w_H$, $w_K$, $w_a$, and $w_i$ in Equation~12 are left to user input. By altering the weight factors, the objective function is biased towards a given property or structure, allowing users to decide what structures, compositions, or features of the energy landscape are most important. 
In this work, a simple data-driven approach was used to optimize the values for $w_H$, $w_K$, and $w_a$ to generate well behaved test-case potentials. 

Within this approach, the validation set consisted of DFT calculations of the QOIs for 17 FCC SQS given by A$_{n}$B$_{32-n}$ with $n \in \{ 1,2,4,\ldots,28,30,31 \}$. And a test matrix of 64 QOI error weight vectors was constructed of the form: 
\begin{equation}
    \label{eq:errorWeightMatrix}
    W_{QOI} = \begin{bmatrix} w_H \\ w_K \\ w_a \end{bmatrix} \textrm{ : for } {w_H, w_K, w_a} \in \{0.8, 1.0, 1.2, 1.4, 1.6\}. 
\end{equation}
RAMPAGE fits were run with each weight factor vector in $W_{QOI}$, and for each of the 64 resulting potentials, the QOIs for the validation set are recomputed using LAMMPS. The distribution of binary potentials are evaluated by the sum of root mean square errors of the QOIs with respect to DFT calculations. For the potentials discussed in Section~4.2, the RMSE vectors associated with each validation set structure were multiplied by an error weight consistent with the logistic function used to weight the training sets to silver and aluminum rich. 

\begin{figure}[htbp]
         \centering
         \includegraphics[width=0.9\textwidth]{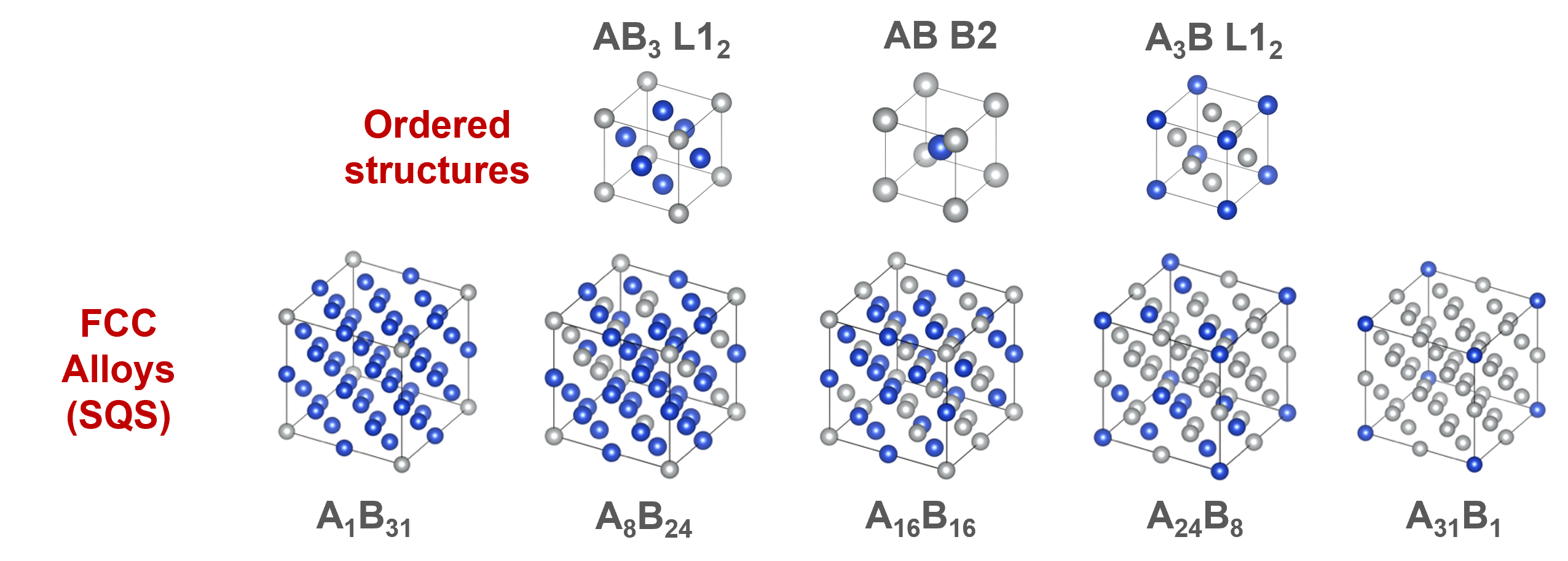}
    \caption{Candidate training set structures consisting of 5 FCC special quasi-random structures and three additional intermetallic structures with L1$_2$ and B2 ordering.
    }
    \label{fig:trainingSet}
\end{figure}

\begin{figure}[htbp]
         \centering
         \includegraphics[width=0.9\textwidth]{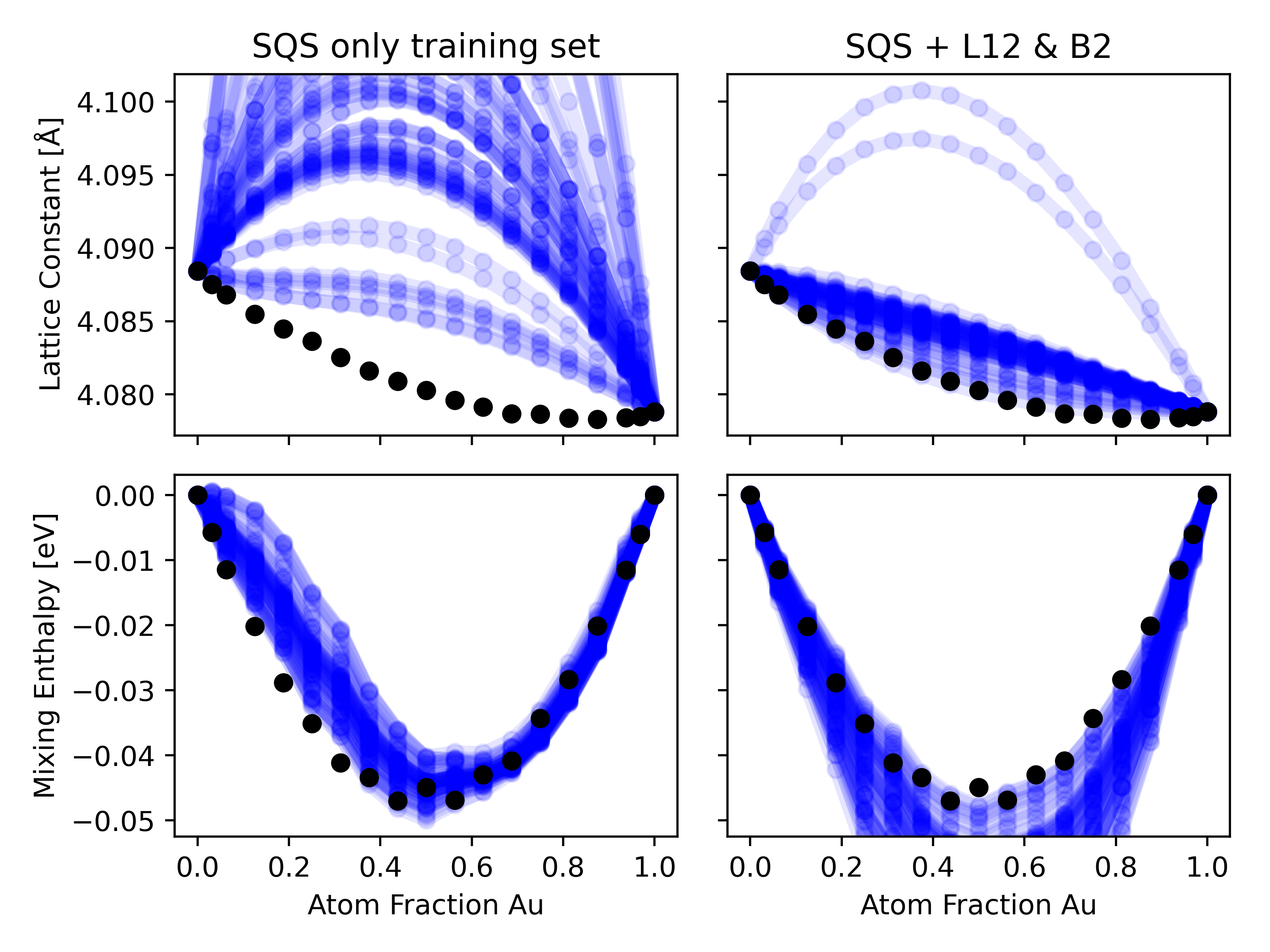}
    \caption{ Ag-Au property calculations for 64 different RAMPAGE fits with DFT computed properties in black. Computed lattice constants (top) and mixing enthalpies (bottom) is compared for fits with training sets that only included only SQS structures (left) and those that included SQS structures and ordered intermetallics (right).}
    \label{fig:errorweighttestAgAu}
\end{figure}

The need for sensible choice of of training set and error weights is indicated in Figure~\ref{fig:errorweighttestAgAu}, which shows the lattice constant and mixing enthalpy results for two preliminary Ag-Au error weight tests. One was produced using a training set consisting of 5 SQSs at composition intervals of 0.03125, 0.25, 0.5, 0.75, and 0.03125 atomic percent aluminum. The other was fitted with the 5 SQSs along with the L1$_{2}$, and B2 intermetallic structures. The Ag-Au system is fully miscible with no stable intermetallics; however, restricting the training set to only alloy structures results in potentials that uniformly over-predict lattice constant and fail to match DFT mixing enthalpy for Ag rich compositions. Including non-ground state intermetallic structures resulted in fits that better captured the DFT data for both properties. 

As no general trend was found to guide the selection of training sets,
the error weight optimization scheme discussed above was applied to several candidate training sets to find the training set that best reproduced the QOIs with the minimum number of structures. The final error weights and training sets for potentials used in the main text are shown in Table~\ref{table:errWeights}

\begin{table}[htbp]
\begin{center}
\begin{tabular}{c | c c c c } 
System & $w_H$ & $w_K$ & $w_a$ & Training Set \\ 
 \hline
 Ag-Al : Ag Focus & 1.6 & 1.2 & 1.6 & 8 SQS structures\\
 Ag-Al : Al Focus & 1.0 & 0.8 & 1.2 & 8 SQS structures\\
 Ag-Au & 1.6 & 1.6 & 0.8 & 5 SQS structures with L1$_2$ and B2 \\
 Ag-Cu & 1.2 & 1.0 & 1.0 & 5 SQS structures with L1$_2$ and B2 \\
 Au-Cu & 0.8 & 1.2 & 0.8 & 5 SQS structures\\
\end{tabular}
\end{center}
\caption{Optimized error weights used for the potentials discussed in Section~4.}
\label{table:errWeights}
\end{table}

\pagebreak

\bibliographystyle{unsrt} 
\bibliography{references}




